\documentclass[aps,reprint,
superscriptaddress,
amsmath, amssymb,
floatfix]{revtex4-2}

\usepackage{graphicx}
\usepackage{subfigure}
\usepackage{dcolumn}
\usepackage{bm}

\usepackage{amsfonts}
\usepackage{color}
\usepackage{array}
\usepackage{booktabs}
\usepackage{multirow}
\usepackage{mathrsfs}

\usepackage{hyperref}
\usepackage{cleveref}
\crefname{figure}{\textcolor{black}{Fig.}}{\textcolor{black}{Fig.}}
\crefname{table}{\textcolor{black}{Tab.}}{\textcolor{black}{Tab.}}
\crefname{equation}{\textcolor{black}{Eq.}}{\textcolor{black}{Eq.}}
\crefname{section}{\textcolor{black}{Sec.}}{\textcolor{black}{Sec.}}

\begin{document}
\title{Prediction of A15 Tilt Grain Boundary Structures}

\author{Wenwen Zou}
\altaffiliation{These authors contribute equally to this work.}

\author{Zihan Su}
\altaffiliation{These authors contribute equally to this work.}

\author{Juan Zhang}

\author{Kai Jiang}
\email{kaijiang@xtu.edu.cn}
\affiliation{Hunan Key Laboratory for Computation and Simulation in Science and Engineering, Key Laboratory of Intelligent Computing and Information Processing of Ministry of Education, School of Mathematics and Computational Science, Xiangtan University, Xiangtan, Hunan, China, 411105}

\begin{abstract}
In this work, we present a theoretical method to predict all coincidence site lattice (CSL) tilt grain boundaries (GBs) in A15, especially high-$\Sigma$ CSL GBs.
This method includes a modified Farey diagram (MFD) and a computational framework based on the 3D phase field crystal model.
Applied to $[001]$ CSL symmetric tilt grain boundaries (STGBs) in A15, this method identifies building blocks of A15 GBs, known as structural units (SUs).
The MFD predicts the quantity and proportion of SUs within GBs.
The developed computational approach further determines the arrangement of these SUs.
The predictive rule reveals the SU arrangement of A15 $[001]$ CSL STGBs.
\end{abstract}

\maketitle

\section{Introduction}
\begin{figure}[!hbpt]
\centering
  \includegraphics[width=0.45\textwidth]{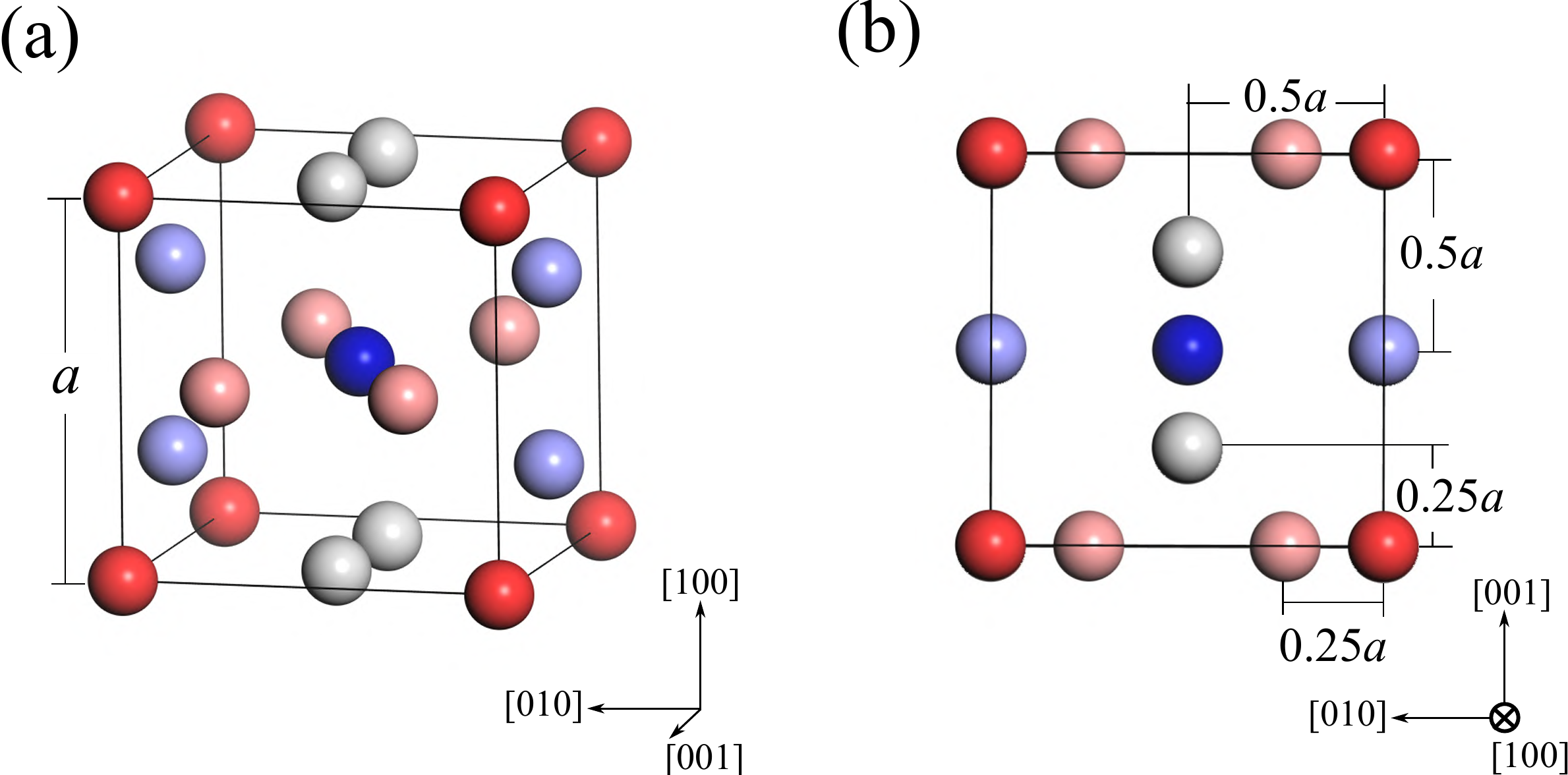}
\caption{A15 structure.
(a) The A15 unit cell contains eight spheres.
Two spheres occupy the corner (red) and the center (blue), constituting a BCC lattice.
The other three pairs (pink, gray, light blue) reside equidistantly on each face along the axes.
The side length of the unit cell is $a$.
(b) The view of the A15 unit cell along the $[100]$ direction.
The spacing between spheres is marked.}
\label{fig:A15_bulk}
\end{figure}

The unit cell of A15, as shown in \cref{fig:A15_bulk}, contains eight spheres.
Two spheres occupy positions in body-centered cubic (BCC) lattice, referred to as lattice spheres.
The remaining three pairs of spheres, distributed on each face in mutually orthogonal chain structures, are termed decorative spheres \cite{ziherl2001maximizing}.
A15 materials such as the Nb$_3$Sn intermetallic compound are superconducting materials with high current density, critical temperature, and critical magnetic field, thus attracting much attention \cite{dew1975superconducting,mao2018ground,qiao2019intrinsic}.
These materials often exist in polycrystalline forms containing grain boundaries (GBs).
The atomic arrangement at GBs significantly influences electron transport and flux pinning behaviors \cite{godeke2005upper,lee2020grain,sitaraman2021effect}.
Therefore, identifying the GB structure is crucial for the design of A15 materials.

Most current studies on GBs in A15 materials focus on GB diffusion \cite{farrell1974grain,sandim2013grain,oh2022diffusion}, GB compositions \cite{suenaga1983chemical}, and conductive effects \cite{sitaraman2021effect}.
A few works have investigated the structures of low-$\Sigma$ coincidence site lattice (CSL) GBs in A15 at the atomic scale, such as $\Sigma 1 ~(110)$, $\Sigma 5~(120)$, and $\Sigma 3~(112)$ GBs \cite{lee2020grain,kelley2020ab,oh2022diffusion}.
However, polycrystalline materials contain numerous high-$\Sigma$ CSL GBs, whose structures remain unknown.
Existing studies lack a systematic theoretical framework for predicting the structures of high-$\Sigma$ CSL GBs and their atomic arrangement rules.
Moreover, experimental characterization methods are limited, making it difficult to directly resolve atomic-scale details of complex interfaces.

\begin{figure*}[!hbpt]
\centering
 \includegraphics[width=0.9\textwidth]{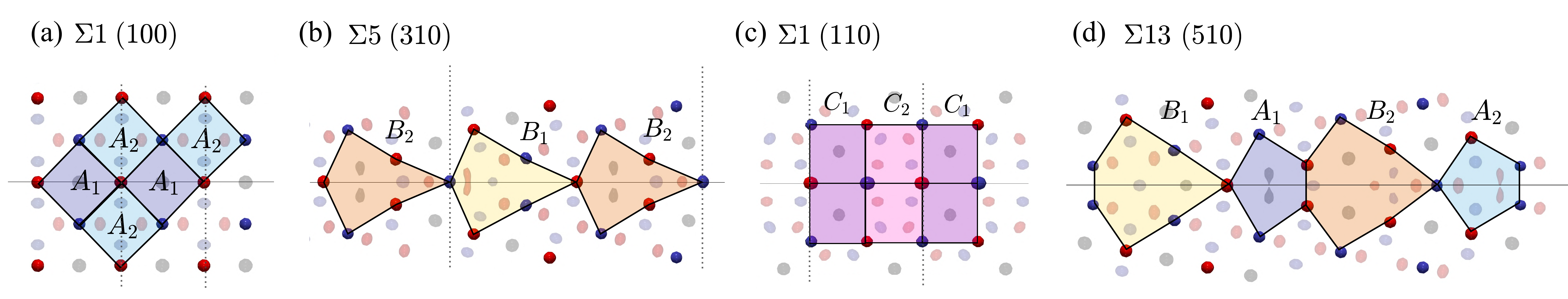}
\caption{Low-$\Sigma$ CSL $[001]$ STGBs in A15 viewed along $[001]$ direction.
(a) $\Sigma 1~(100)~(0^{\circ})$ GB.
(b) $\Sigma 5~(310)~(36.9^{\circ})$ GB.
(c) $\Sigma 1~(110)~(90^{\circ})$ GB.
(d) $\Sigma 13~(510)~(22.6^{\circ})$ GB.
SUs are labeled $A_1$ (dark blue), $A_2$ (light blue), $B_1$ (yellow), $B_2$ (orange), $C_1$ (purple), $C_2$ (pink).
In these GBs, SUs arrange periodically as $(A_1A_2)$, $(B_1B_2)$, $(C_1C_2)$, and $(B_1A_1B_2A_2)$, respectively.
}
\label{fig:A15}
\end{figure*}

To fill the gap, this work proposes a structural prediction method for all CSL GBs by developing a modified Farey diagram (MFD) and a computational framework based on the 3D phase field crystal model \cite{jiang2022tilt,jiang2024numerical,GBsoftware1,GBsoftware2}.
We systematically study $[001]$ CSL symmetric tilt grain boundaries (STGBs) in A15 covering misorientation angles from $0^\circ$ to $90^\circ$.
By combining the structural unit model with numerical computations, we identify three categories of structural units (SUs), and find three delimiting GBs: $\Sigma 1~(100)~(0^\circ)$, $\Sigma 5~(310)~(36.9^\circ)$, and $\Sigma 1~(110)~(90^\circ)$ GBs.
We divide misorientation angles of $[001]$ CSL GBs into two intervals: $0^\circ$-$36.9^\circ$ and $36.9^\circ$-$90^\circ$, using the MFD to predict arrangement rules of GB SUs in each interval.
Finally, we apply the computational framework to determine specific arrangements of GB SUs, establishing predictive rules for A15 GB structures.

\section{Theoretical framework}

\subsection{Modified Farey diagram in A15 GBs}

In this section, we introduce the structural unit model and the MFD for A15 $[001]$ CSL STGBs.
Based on SUs and delimiting GBs, we establish a hierarchical relationship between misorientation angles and SU arrangements.
By adjusting the initial fraction in the classical Farey diagram, an MFD is developed to enable systematic prediction of SU compositions in GBs.

\subsubsection{Structural unit model}

The structural unit model has been used to study structural characteristics of CSL GBs \cite{sutton1983structure,wang1984computer,li2016grain,han2017grain}, and it contains two key concepts.

(i) SU representation \cite{bishop1968coincidence, sutton1983structure}.
A GB is a combination of building blocks, where SUs represent their two-dimensional projections.
\cref{fig:A15} shows four low-$\Sigma$ $[001]$ CSL STGBs in A15 viewed along the $[001]$ direction.
The SU shape can be determined by lattice spheres, i.e., BCC lattice.
SUs of A15 GBs include type $A$ (diamond), type $B$ (kite-shaped), and type $C$ (rectangular).
These SUs are further subdivided into $A_1$, $A_2$, $B_1$, $B_2$, $C_1$, and $C_2$ based on the colors of their lattice spheres.
For instance, the lattice sphere colors of $A_1$ and $A_2$ are interchanged (blue $\leftrightarrow$ red).
The SU periodic arrangements of $\Sigma 1~(100)$, $\Sigma 5~(310)$, $\Sigma 1~(110)$, and $\Sigma 13~(510)$ GBs are $(A_1A_2)$, $(B_1B_2)$, $(C_1C_2)$ and $(B_1A_1B_2A_2)$, respectively.

(ii) SU combination \cite{sutton1983structure}.
A GB composed of a single type of SU is a delimiting GB \cite{bristowe1985structural}.
For A15, $\Sigma 1~(100)~(0^\circ)$, $\Sigma 5~(310)~(36.9^\circ)$, and $\Sigma 1~(110)~(90^\circ)$ GBs are delimiting GBs, each containing only type $A$, type $B$, and type $C$ SUs, respectively.
A GB with misorientation $\theta$ between the angles ($\theta_1$ and $\theta_2$) of two adjacent delimiting GBs consists of their SUs in a ratio of
\begin{equation}
\frac{n_1}{n_2} = \frac{l_2}{l_1} \frac{\sin [(\theta_{2} - \theta)/2]}{\sin [(\theta - \theta_{1})/2]}.
\label{eq:rate1}
\end{equation}
Here, $n_1$ and $n_2$ represent the numbers of SUs from two delimiting GBs.
$l_1$ and $l_2$ are lengths of SUs along the GB plane in these two delimiting GBs.
As an example, \cref{fig:A15}(d) shows that $\Sigma 13~(510)$ GB with misorientation $22.6^\circ \in (0^\circ, 36.9^\circ)$ consists of SUs $A$ and $B$ in a $1:1$ ratio according to \cref{eq:rate1}.

\subsubsection{Modified Farey diagram}

\begin{figure}[!hbpt]
\centering
    \includegraphics[width=0.24\textwidth]{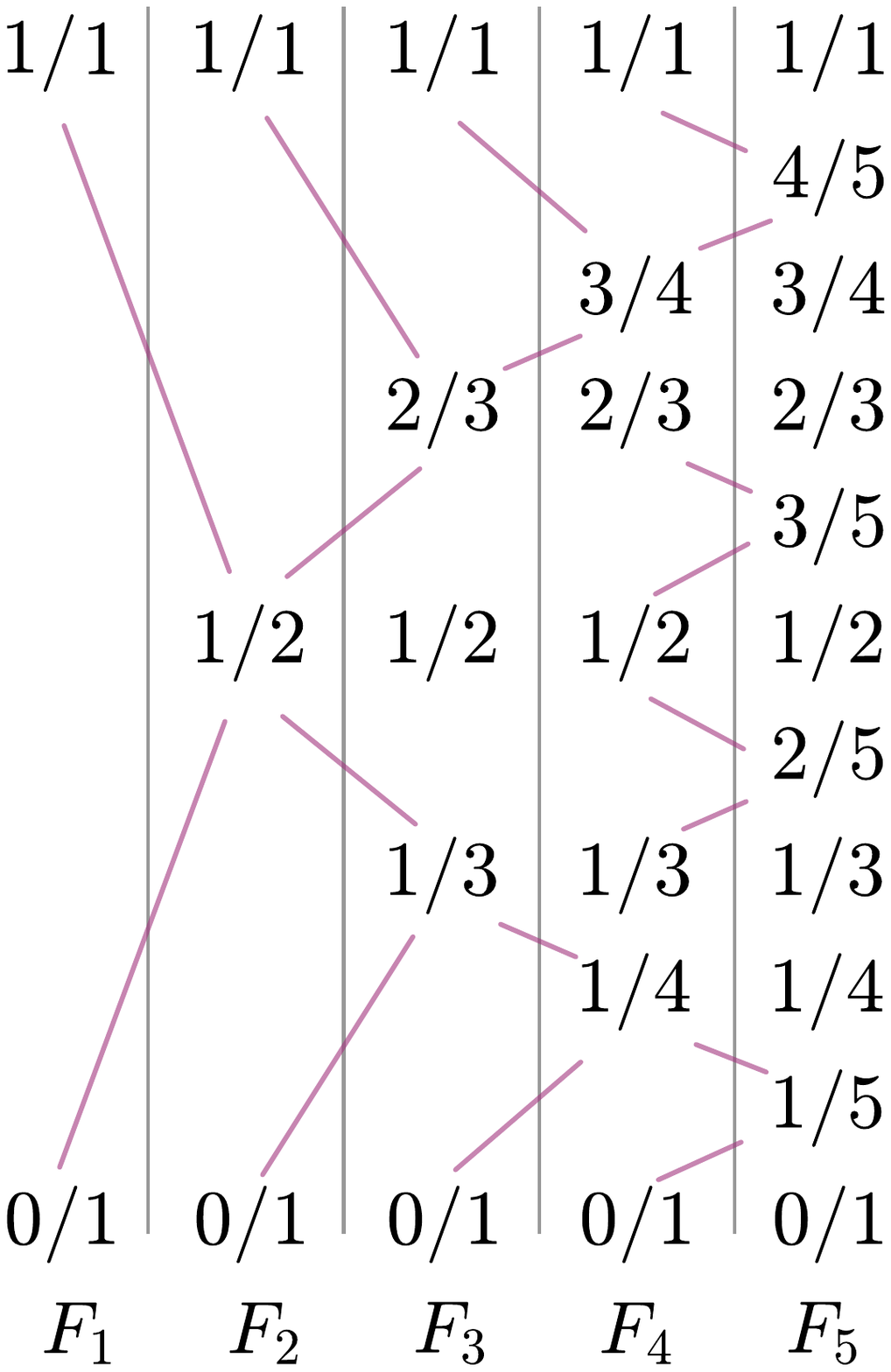}
\caption{Classical Farey diagram.
It is a binary tree built from Farey sequences.
The Farey sequence $F_N$ is an ordered set of irreducible fractions with denominators not exceeding $N$.
The initial fractions are $0/1$ and $1/1$, and new node fractions are generated by the Farey summation in \cref{eq:sum}.}
\label{fig:FareyDia}
\end{figure}

\begin{figure}[!hbpt]
\centering
    \includegraphics[width=0.50\textwidth]{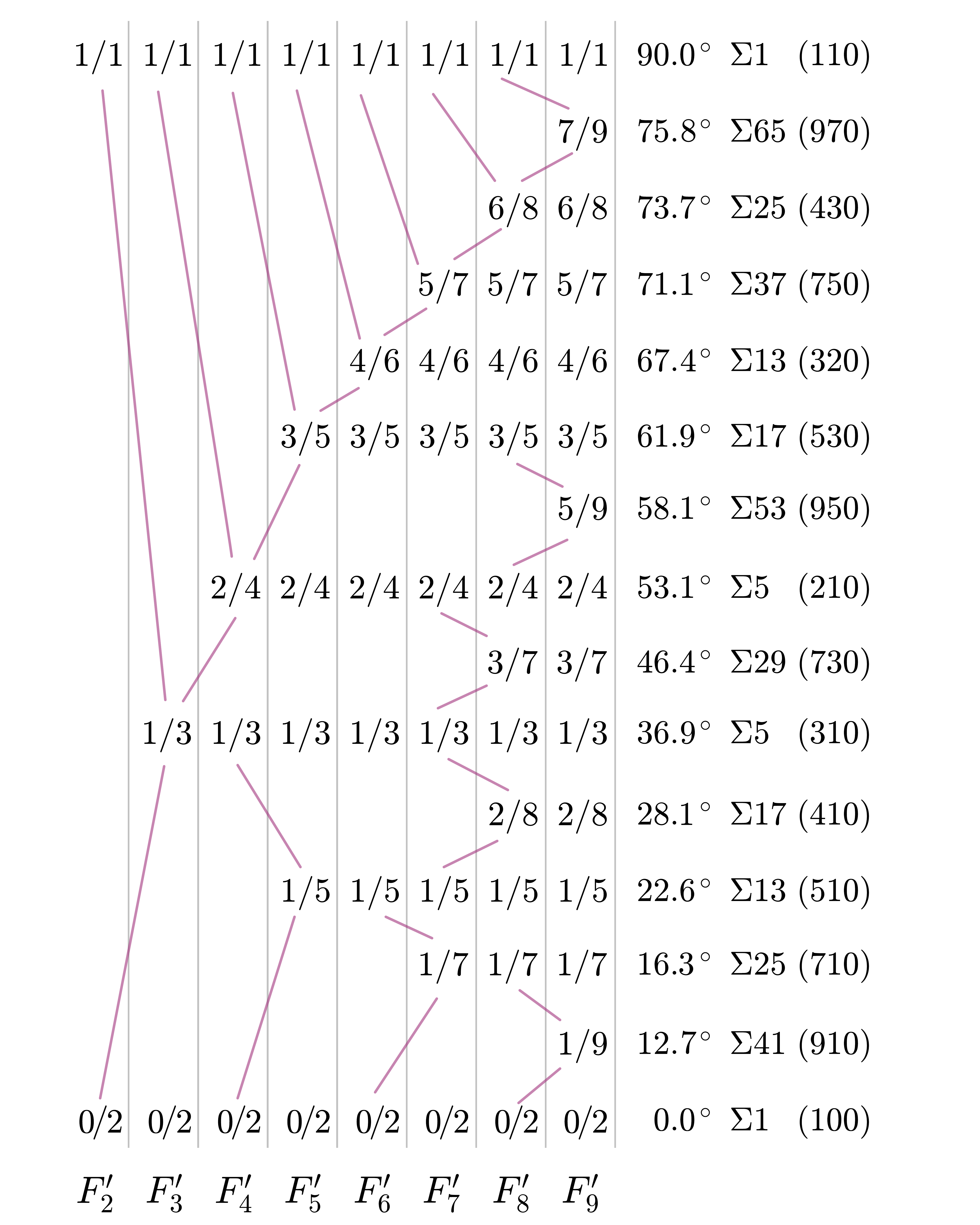}
\caption{MFD illustrating the hierarchical structure of $[001]$ CSL GBs.
It alters the initial fraction of the classic version from $0/1$ to $0/2$.
New node fractions are generated by Farey summation in \cref{eq:sum} and may be reducible.
The fraction $q/p$ maps to a GB with misorientation angle $ 2\arctan(q/p)$, annotated on the right side of the diagram.}
\label{fig:farey}
\end{figure}

The Farey diagram was originally proposed to describe the distribution of rational numbers \cite{hardy1979introduction}.
As shown in \cref{fig:FareyDia}, the classical Farey diagram is a binary tree built by iteratively generating Farey sequences.
The initial fractions of the classic Farey diagram are $0/1$ and $1/1$.
The node between two adjacent fractions is derived through Farey summation
\begin{equation}
\label{eq:sum}
\dfrac{q_1}{p_1} \boxplus \dfrac{q_2}{p_2} = \dfrac{q}{p},
\end{equation}
where $p = p_1 + p_2$ and $q = q_1 + q_2$.
Classical Farey fractions are all irreducible.
The $N$-th generation Farey sequence $F_N$ is an increasing sequence of irreducible fractions with denominators not exceeding $N$.
For example, the Farey sequence $F_3$ is $\{0/1, 1/3, 1/2, 2/3, 1/1\}$.

Farey diagram has recently been extended to GBs to describe the relationship between SU arrangements in CSL GBs and the distribution of rational numbers  \cite{inoue20213d,inoue2021arrangement}.
Here, we develop a MFD to depict the hierarchy of A15 $[001]$ CSL STGBs, and further predict arrangements of GB SUs, as shown in \cref{fig:farey}.
Lattice spheres form a BCC Bravais lattice with fourfold rotational symmetry, thus we investigate GBs with misorientation angles between $0^\circ$ and $90^\circ$.
The MFD changes $0/1$ in the classical Farey diagram to $0/2$.
The Farey fraction $0/2$ corresponds to $\Sigma 1~(100)$ GB.
This modification ensures that the first two Farey sequences correspond to delimiting GBs.
The modified Farey sequence is denoted as $F^\prime_N$, where $N \geq 2$.
For example, $F^\prime_3=\{0/2, 1/3, 1/1\}$ and $F^\prime_4=\{0/2, 1/3, 2/4, 1/1\}$.
Unlike the classical Farey fractions, modified Farey fractions may be reducible.
The misorientation angle of the GB corresponding to Farey fraction $q/p$ is $\theta = 2 \arctan{(q/p)}$.
We divide misorientation angles $\theta$ from $0^\circ$ to $90^\circ$ into two ranges ($0^\circ$-$36.9^\circ$ and $36.9^\circ$-$90^\circ$) based on delimiting GBs.
For $\theta \in (0^{\circ}, 36.9^{\circ})$ ($0 < q/p < 1/3$), the GB consists of SUs $A$ and $B$; for $\theta \in (36.9^{\circ}, 90^{\circ})$ ($1/3 < q/p < 1$), it comprises SUs $B$ and $C$.
\cref{eq:sum} indicates that $(pq0)$ GB is composed of SUs from $(p_1 q_10)$ and $(p_2q_20)$ GBs, with the number of symbols in former's SU periodic sequence equaling sum of latter two.
Taking $0/2 \boxplus 1/3 = 1/5$ as an example, the SU periodic sequence $(B_1A_1B_2A_2)$ of $\Sigma 13~(510)$ GB consists of $(A_1A_2)$ from $\Sigma 1~(100)$ GB and $(B_1B_2)$ from $\Sigma 5~(310)$ GB.
The proportion and quantity of various SUs in SU periodic sequences of CSL GBs could be predicted by the positions of Farey fractions.

\subsection{Computational framework for GBs}
\label{subsec:framework}

Now, we present a computational framework capable of accurately calculating $[001]$ CSL STGBs in A15, particularly high-$\Sigma$ CSL GBs.
We use phase field crystal model \cite{elder2002modeling, brazovskii1975phase} to calculate GBs, with the free energy functional
\begin{equation}\label{eq:PFCmodelforGBs}
\begin{aligned}
	E\left[ \psi(\bm{r}) \right] = \frac{1}{V(\Omega)} \int_\Omega \, & \left\lbrace \frac{1}{2}  \left[ \left( 1 + \nabla^2 \right)\psi \right]^2 \right.\\
	& \left. + \frac{\tau}{2!} \psi^2 -\frac{\gamma}{3!} \psi^{3}+\frac{1}{4!} \psi^{4} \right\rbrace d \bm{r},
\end{aligned}
\end{equation}
where $\psi(\bm{r})(\bm{r}=(x,y,z)^T)$ is the local 3D atomic density.
$\Omega$ is a region in $\mathbb{R}^3$, whose volume is $V(\Omega)$.
Mass conservation requires
\begin{equation}\label{eq:mass_constraint}
\begin{aligned}
    \frac{1}{V(\Omega)}\int_\Omega \, \psi(\bm{r})\, d \bm{r}
    =0.
\end{aligned}
\end{equation}
The time evolution follows a conserved dynamics equation
\begin{equation}\label{eq:time_evolution}
\begin{split}
	\frac{\partial \psi}{\partial t} &= -\frac{\delta E}{\delta \psi},\\
	\frac{\delta E}{\delta \psi} &= \left( 1 + \nabla^2 \right)^2 \psi + \mathcal{P}(\tau \psi  - \frac{\gamma}{2}\psi^2 + \frac{1}{6}\psi^3),
\end{split}
\end{equation}
where $\mathcal{P}\varphi= \varphi - \frac{1}{V(\Omega)}\int_\Omega \, \varphi(\bm{r})d \bm{r}$ serves as a projection operator imposing the mass conservation constraint.

\begin{figure}[!hbpt]
\centering
 \includegraphics[width=0.45\textwidth]{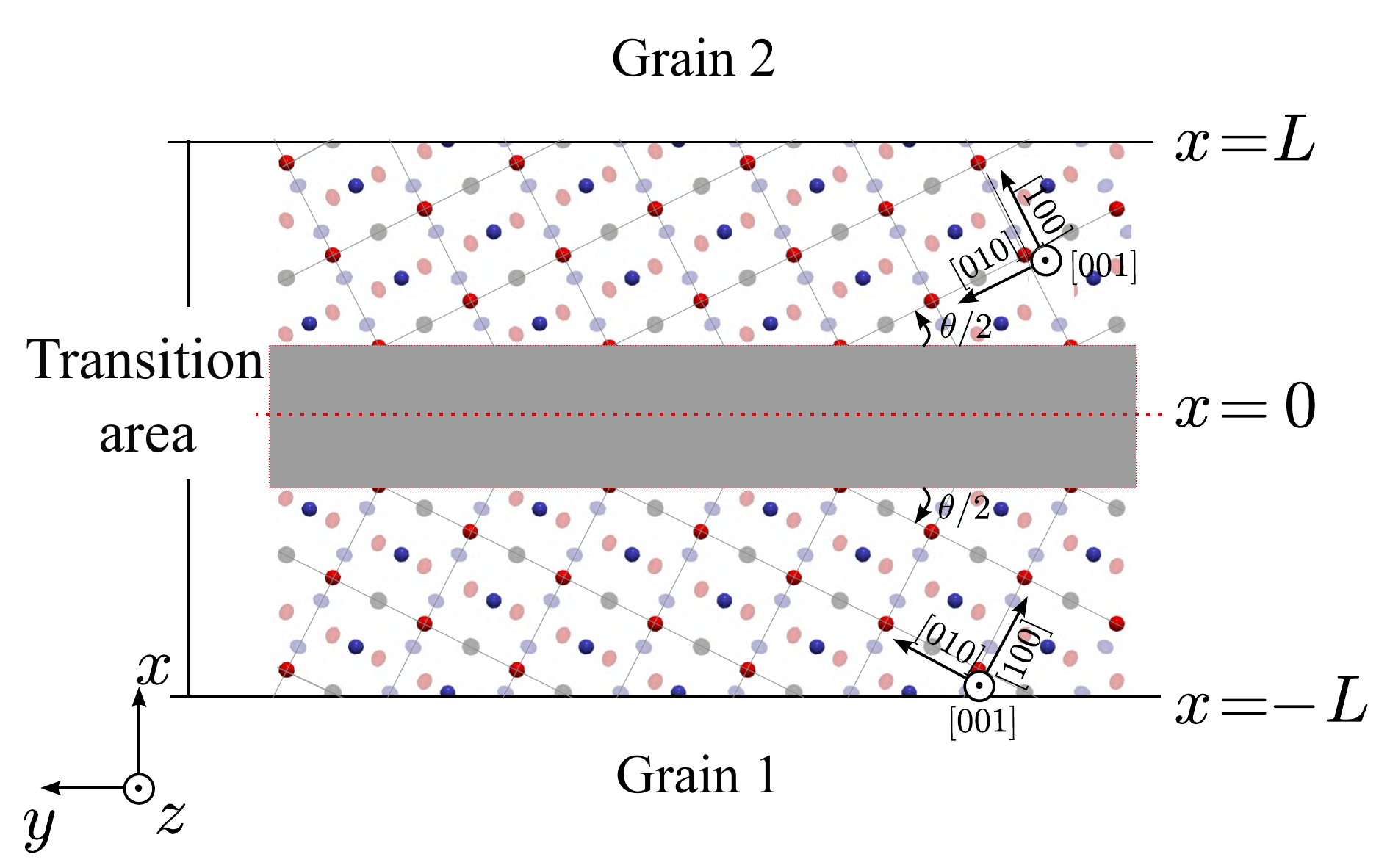}
\caption{Schematic of $[001]$ STGBs in A15.
Planes $x = -L$ and $x = L$ divide the 3D space into regions.
Grain 1 (Grain 2) rotates clockwise (counterclockwise) by $\theta/2$ around $[001]$ direction, occupying the half-space $x\leq -L$ ($x\geq L$).
A transition region containing GBs lies between them.
The coordinate system and crystal orientations are labeled.
}
\label{fig:GB}
\end{figure}

The GB simulation employs a 3D phase field crystal model, with the schematic shown in \cref{fig:GB}.
Planes $x = \pm L$ partition the 3D space.
Grain 1 (Grain 2), located in the half-space $x\leq -L$ ($x\geq L$), undergoes a clockwise (counterclockwise) rotation of $\theta/2$ about $[001]$ axis.
Here, $\theta$ is the misorientation angle between the two grains.
The transition zone between the two planes contains GBs accommodating structural relaxation.
The GB structure is periodic along $y$ and $z$ directions.
In the $x$ direction, nonhomogeneous Dirichlet boundary conditions are imposed, while A15 crystal structure is maintained at $x = \pm L$.

Next, we establish the discrete scheme of \cref{eq:time_evolution}.
The model parameters $\tau = 0.0$ and $\gamma = 1.3$ are selected according to the stable A15 region of the phase diagram \cite{mcclenagan2019landau}.
The A15 structure can be represented by a Fourier series,
\begin{equation}
	\psi_{0}(\bm{r})=\sum_{\bm{h} \in \mathbb{Z}^{3}} \hat{\psi}_{0}(\bm{h}) \exp\left[ i(\bm{G}_0 \bm{h})^{T} \bm{r} \right],
\end{equation}
where the $3 \times 3$ matrix $\bm{G}_0 = b\bm{I}$ represents the reciprocal lattice, with $b$ being the length of reciprocal lattice vectors.
Driven by the minimization of the free energy functional, alternating iterations of $\bm{G}_0$ and $\hat{\psi}_{0}(\bm{h})$ are performed to obtain a stable A15 structure \cite{jiang2014numerical}.
The optimal side length of the A15 unit cell is $ a = 2\pi/b = 14.7 $.
The $[100]$, $[010]$, and $[001]$ directions of A15 correspond to the $x$, $y$, and $z$ axes of the coordinate system.
Rotating clockwise (counterclockwise) around $[001]$ by $\theta/2$ generates grain 1 (grain 2), denoted as $\psi_s(s=1,2)$,
\begin{equation*}\label{eq:RotatedGrain}
	\begin{split}   
		\psi_{s}(\bm{r}) &= \psi_{0}\left(R_{s}^T \bm{r}\right) \\
		&= \sum_{\bm{h} \in \mathbb{Z}^{3}} \hat{\psi}_{0}(\bm{h}) \exp\left[i \left(R_{sx} \bm{G}_0 \bm{h}\right) x \right]\exp\left[i \left(\tilde{R}_{s} \bm{G}_0 \bm{h}\right)^{T} \tilde{\bm{r}} \right].
	\end{split}
\end{equation*}
Here, the rotation matrix is $R_s = \begin{bmatrix} \cos(\theta_s) & -\sin(\theta_s) & 0 \\ \sin(\theta_s) & \cos(\theta_s) & 0 \\ 0 & 0 & 1 \end{bmatrix}$ with $ \theta_1 = -\theta/2 $ and $ \theta_2 = \theta/2 $.
CSL GBs require $\tan(\theta/2)$ to be a rational number \cite{zou2025quasiperiodic}.
$R_{sx}$ and $\tilde{R}_s$ are the first row and the last two rows of $R_s$ respectively, and $\tilde{\bm{r}}=(y, z)^T$.
The terms $\hat{\psi}_{0}(\bm{h}) \exp\left[i \left(R_{sx} \bm{G}_0 \bm{h}\right) x \right]$ and $\tilde{R}_{s} \bm{G}_0$ are rewritten as $\hat{\psi}_{s}(x, \bm{h})$ and $\bm{G}_{s}$, respectively. 
Therefore, $\psi_{s}(\bm{r})$ can be simplified to
\begin{equation}\label{eq:RotatedGrain1}
	\begin{split}   
		\psi_{s}(x, \tilde{\bm{r}})= \sum_{\bm{h} \in \mathbb{Z}^{3}} \hat{\psi}_{s}(x, \bm{h}) \exp\left[i\left(\bm{G}_{s} \bm{h}\right)^{T} \tilde{\bm{r}}\right].
	\end{split}
\end{equation}

Since CSL GBs are periodic in the $y$-$z$ plane, we can find a common periodic function space,
\begin{equation}\label{eq:common_space}
\begin{aligned}
    W = \mbox{span} \left\lbrace \exp\left[i\left(\bm{G} \bm{h}\right)^{T} \tilde{\bm{r}}\right], ~\bm{h} \in \mathbb{Z}^{2} \right\rbrace.
\end{aligned}
\end{equation}
This function space is composed of linear combinations of 2D periodic exponential functions.
Grains and GB can be uniformly represented within the function space $W$.
The two grains in \cref{eq:RotatedGrain1} are re-expressed as
\begin{equation}
\label{eq:grains_in_common_space}
	\begin{split} 
		\psi_{s}(x, \tilde{\bm{r}}) =\sum_{\bm{h} \in \mathbb{Z}^{2}} \hat{\psi}_{s}(x, \bm{h}) \exp\left[i\left(\bm{G} \bm{h}\right)^{T} \tilde{\bm{r}}\right].
	\end{split}
\end{equation}
Boundary conditions along $x$ direction of the system are defined by matching the values and normal derivatives of the grains at $x = \pm L$,
\begin{equation}
\label{eq:xBC}
\begin{aligned}
    \frac{\partial^{m} \psi(x, \tilde{\bm{r}})}{\partial x^{m}}\Bigg|_{x=-L} & =\frac{\partial^{m} \psi_{1}(x, \tilde{\bm{r}})}{\partial x^{m}}\Bigg|_{x=-L},\\ 
    \frac{\partial^{m} \psi(x, \tilde{r})}{\partial x^{m}}\Bigg|_{x=L} & =\frac{\partial^{m} \psi_{2}(x, \tilde{\bm{r}})}{\partial x^{m}}\Bigg|_{x=L},
\end{aligned}
\end{equation}
for $m=0$, $1$.
The initial value of the system can be constructed by connecting two grains with a smooth function,
\begin{equation}
    \label{eq:initial_state}
	\begin{aligned}
		\hat{\psi}^{(0)}(x, \bm{h}) = \left[ 1-b(x) \right] \hat{\psi}_1(x, \bm{h}) + b(x) \hat{\psi}_2(x, \bm{h}),
	\end{aligned}
\end{equation}
where $b(x) = \left[1-\tanh (\sigma x)\right]/2$ with $\sigma = 0.1$, ensuring that $b(-L)=0$ and $b(L)=1$ to satisfy the boundary conditions in \cref{eq:xBC}.

The temporal discretization of \cref{eq:time_evolution} employs a semi-implicit scheme \cite{jiang2014numerical}, with time step size $\Delta t$.
\begin{equation}
    \label{eq:discrete_equation}
	\begin{split}   
        &\left[ \partial_x^4 - \left( 2\bm{k}^2 -2 \right) \partial_x^2 + \bm{k}^4 - 2\bm{k}^2 + 1 + \frac{1}{\Delta t} \right]\hat{\psi}^{(n+1)}(x,\bm{h}) \\
        = &\left( \frac{1}{\Delta t} - \tau \right)\hat{\psi}^{(n)}(x,\bm{h}) + \frac{\gamma}{2}(\widehat{\psi^2})^{(n)}(x,\bm{h}) - \frac{1}{6}(\widehat{\psi^3})^{(n)}(x,\bm{h}),
	\end{split}
\end{equation}
where $\bm{k}= \bm{G} \bm{h}$ is the reciprocal vector.
These terms $(\widehat{\psi^j})(x,\bm{h}) = \int d \tilde{\bm{r}} \psi^j(x, \tilde{\bm{r}}) \exp\left[-i \left(\bm{G} \bm{h} \right)^T \tilde{\bm{r}}\right], j=2,3$ can be efficiently calculated by the pseudospectral method \cite{zhang2008efficient}.
We use generalized Jacobi polynomials to discretize the variable $x$ in \cref{eq:discrete_equation} \cite{cao2021computing}.
The generalized computational framework for arbitrary interfaces can be found in \cite{jiang2022tilt,jiang2024numerical}.
It has been preliminarily applied to calculate CSL and non-CSL GBs for hexagonal and face-centered cubic crystalline materials \cite{jiang2022tilt,zou2025quasiperiodic}.

\section{Results and discussion}
\label{sec:results}

In this section, we combine the MFD with the computational framework to predict GB structures in two misorientation intervals.

\subsubsection{Predicting GBs for $\theta \in (0^\circ,36.9^\circ)$}

\begin{figure*}[!hbpt]
\centering
 \includegraphics[width=0.8\textwidth]{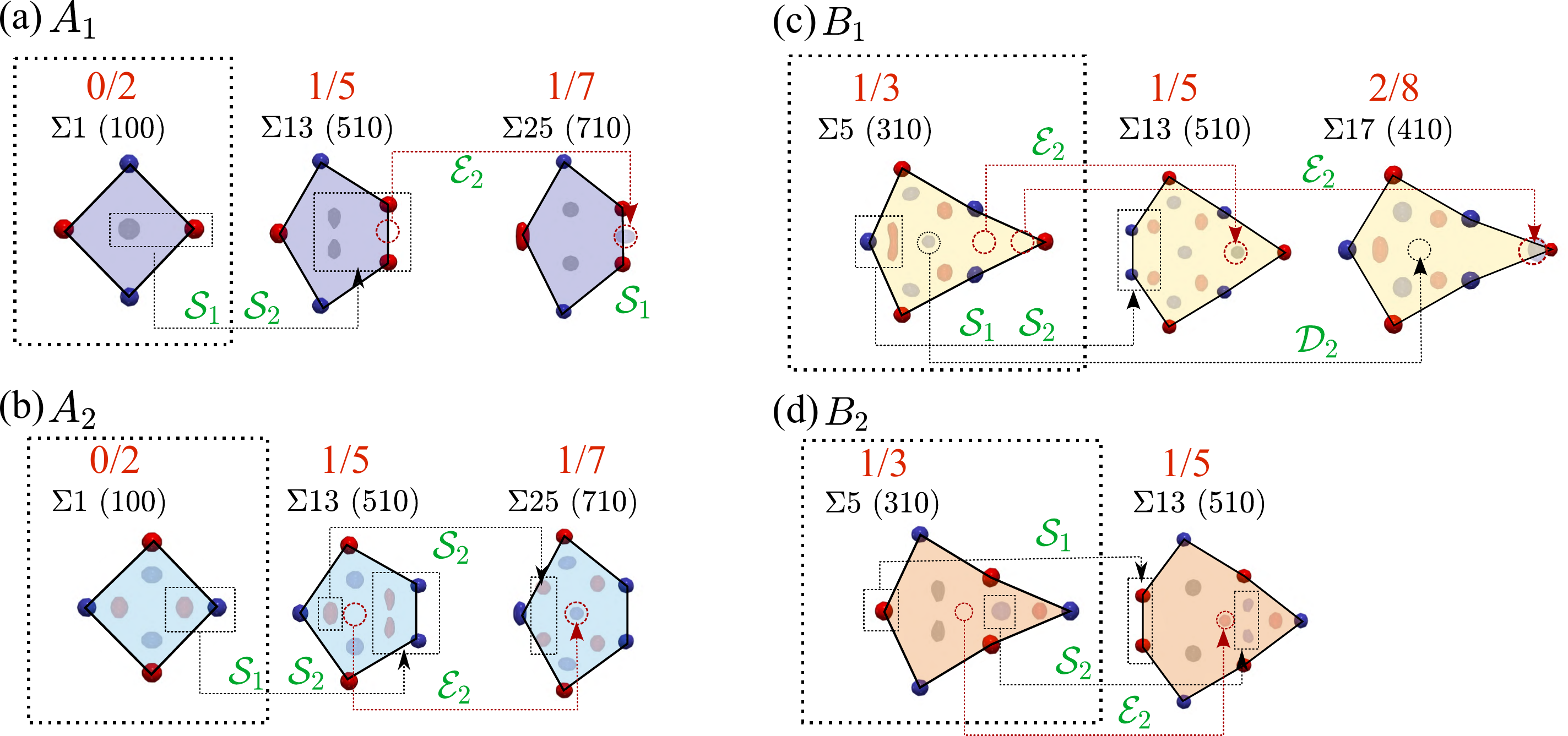}
\caption{
Shapes and deformations of SUs $A_{1}$, $A_{2}$, $B_{1}$, and $B_{2}$ in GBs with $0^{\circ}$-$36.9^{\circ}$ misorientation angles within MFD in \cref{fig:farey}.
SUs are categorized by type and ordered by ascending Farey denominators.
If the same SU appears in multiple GBs, we retain the SU instance with the smallest Farey denominator and remove duplicates.
Dashed boxes indicate basic SUs in delimiting GBs.
Green symbols denote deformation types, where subscript ``1" represents lattice spheres and ``2" represents decorative spheres.
SU deformations include splitting of lattice spheres ($\mathcal{S}_{1}$), splitting of decorative spheres ($\mathcal{S}_{2}$), emergence of new decorative spheres ($\mathcal{E}_{2}$), and disappearance of decorative spheres ($\mathcal{D}_{2}$).
}
\label{fig:SU1}
\end{figure*}

\begin{figure*}[!hbpt]
\centering
 \includegraphics[width=0.7\textwidth]{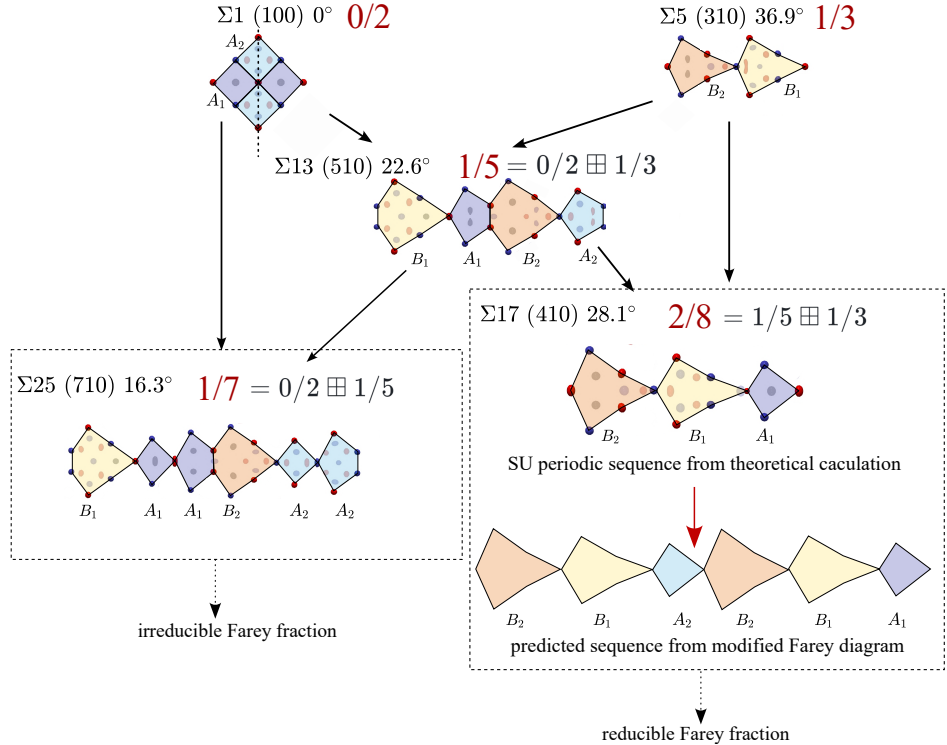}
\caption{
Structural prediction of $[001]$ CSL STGBs in A15 with misorientation angles $0^\circ$-$36.9^\circ$ by MFD and theoretical calculations.
SUs of $\Sigma 13~(510)$ GB are periodically arranged as $(B_1A_1B_2A_2)$.
$\Sigma 25~(710)$ GB (irreducible fraction $1/7$) exhibits a periodic SU sequence $(B_1A_1A_1B_2A_2A_2)$, consistent with MFD prediction as it comprises SUs from both $\Sigma 1~(100)$ and $\Sigma 13~(510)$ GBs.
The periodic sequence of SUs in $\Sigma 17~(410)$ GB (reducible fraction $2/8$) is $(B_2B_1A_1)$, which constitutes half of the sequence predicted by MFD.
}
\label{fig:farey1}
\end{figure*}

GBs with misorientation angles $\theta \in (0^{\circ}, 36.9^{\circ})$ contain type $A$ and $B$ SUs, corresponding to Farey fractions between $0$ and $1/3$.
\cref{fig:SU1} displays concrete shapes of SUs $A_1$, $A_2$, $B_1$, and $B_2$ in delimiting GBs and several specific CSL GBs.
Dashed boxes contain basic shapes of each SU in delimiting GBs.
As the denominator of Farey fraction increases (the generation of Farey sequence rises), SUs undergo deformations, including splitting ($\mathcal{S}$), emergence ($\mathcal{E}$), and disappearance ($\mathcal{D}$).
We use subscripts ``1" and ``2" to distinguish deformations of lattice spheres and decorative spheres.
A lattice sphere splits into two, labeled $\mathcal{S}_1$.
Transformations of decorative spheres involve splitting into two ($\mathcal{S}_2$), new decorative spheres emerging ($\mathcal{E}_2 $), and disappearing ($\mathcal{D}_2$).
SU $A_1$ evolves from $0/2$ to $1/5$ via $\mathcal{S}_1\mathcal{S}_2$, then to $1/7$ through $\mathcal{E}_2$.
We obtain the deformation process of $A_1$, which is $0/2\xrightarrow[  ]{\mathcal{S}_1\mathcal{S}_2}1/5\xrightarrow[]{\mathcal{E}_2}1/7$.
Similarly, the deformation process of $A_2$ is $0/2\xrightarrow[ ]{\mathcal{S}_1\mathcal{S}_2}1/5\xrightarrow[]{\mathcal{S}_2\mathcal{E}_2}1/7$, that of $B_1$ is $1/3\xrightarrow[]{\mathcal{S}_1\mathcal{S}_2\mathcal{E}_2}1/5$ and $1/3\xrightarrow[]{\mathcal{D}_2\mathcal{E}_2}2/8$, and that of $B_2$ is $1/3\xrightarrow[]{\mathcal{S}_1\mathcal{S}_2\mathcal{E}_2}1/5$.
With increasing Farey fraction denominator (i.e., higher sequence generation), the deformation of SUs becomes more complex compared with shapes of basic SUs in delimiting GBs.
SUs derived from the same basic SU retain their category.

Next, we propose the SU arrangement rule in GBs based on MFD.
\cref{fig:farey1} displays SU arrangement within one period for five GBs corresponding to Farey fractions between $0$ and $1/3$ in $F^\prime_8$.
Except for delimiting GBs, SU periodic sequences in other GBs are composed of combinations of $(B_1 A_1 \dots A_1)$ and $(B_2 A_2 \dots A_2)$, where the number of $A_2$ SUs may be zero.
We compare the predicted sequences from MFD with GBs calculated theoretically, classifying Farey fractions as irreducible or reducible.
When Farey fraction is irreducible, we can directly obtain SU periodic sequence for corresponding GB.
For example, the Farey fraction $1/7=0/2\boxplus1/5$ implies that SU sequence ($B_1A_1A_1B_2A_2A_2$) in $\Sigma 25~(710)$ GB combines SU sequences ($A_1A_2$) in $\Sigma 1~(100)$ GB and ($B_1A_1B_2A_2$) in $\Sigma 13~(510)$ GB.
When Farey fraction is reducible, the MFD-predicted sequence requires reduction to match SU periodic sequence.
For instance, $\Sigma 17~(410)$ GB, with Farey fraction $2/8 = 1/3 \boxplus 1/5$, has predicted sequence $(B_2B_1A_1B_1B_2A_2)$, combining $(B_1B_2)$ in $\Sigma 5~(310)$ GB and $(B_1A_1B_2A_2)$ in $\Sigma 13~(510)$ GB.
However, SU periodic sequence of theoretically calculated $\Sigma 17~(410)$ GB is $(B_2B_1A_1)$, which is half of the predicted sequence.
Comparing sequences $(B_2B_1A_1)$ and $(B_1B_2A_2)$, we hypothesize that GBs retain the subsequence with more $A_1$ SUs.

\begin{figure*}[!hbpt]
\centering
 \includegraphics[width=0.7\textwidth]{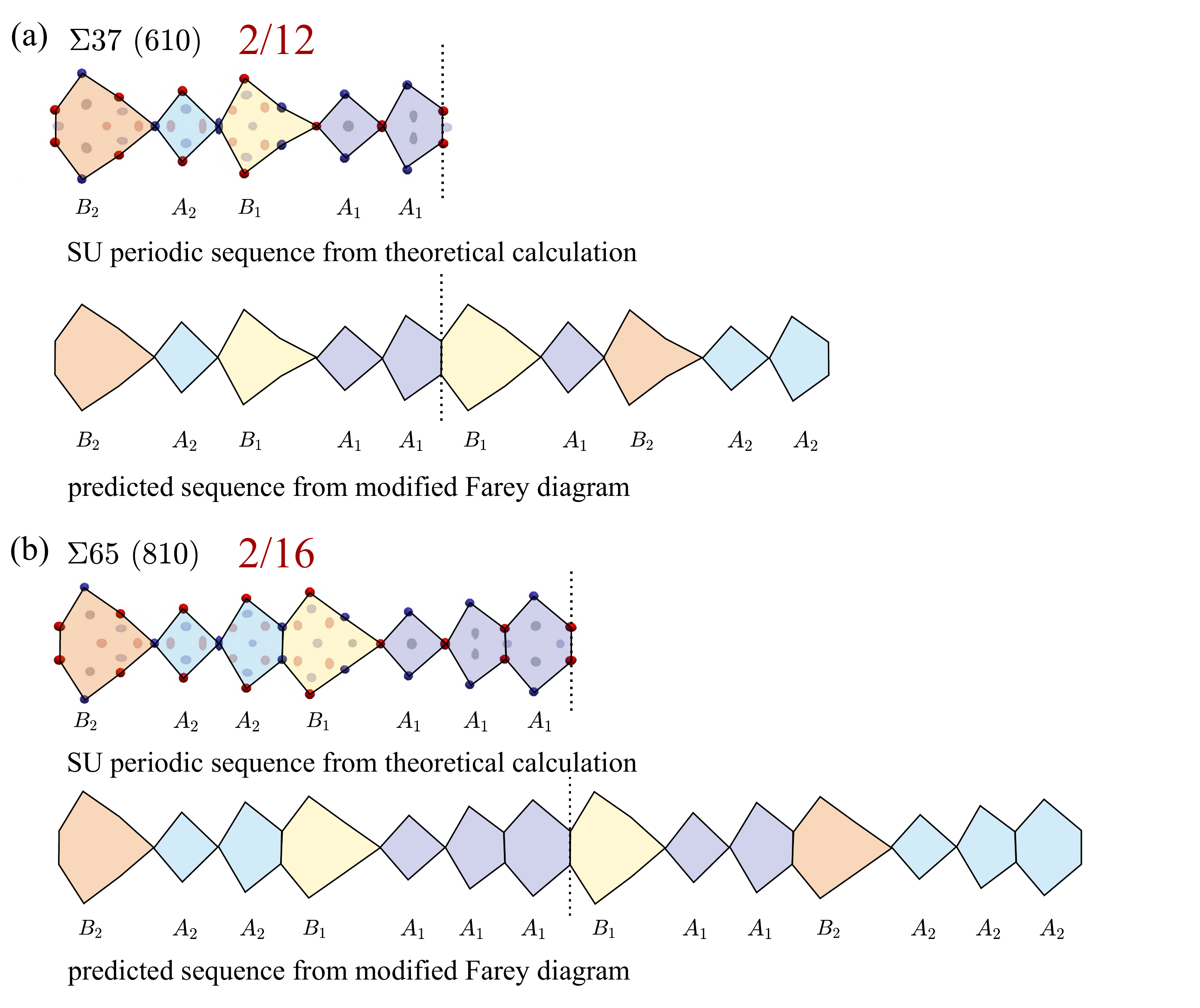}
\caption{SU arrangements in high-$\Sigma$ CSL GBs corresponding to reducible Farey fractions.
(a) For $\Sigma 37 ~(610)$ GB (Farey fraction $2/12$), the theoretical SU sequence is $(B_2A_2B_1A_1A_1)$, while the MFD predicts $(B_2A_2B_1A_1A_1|B_1A_1B_2A_2A_2)$.
(b) For $\Sigma 65 ~(810)$ GB (Farey fraction $2/16$), the theoretical sequence $(B_2A_2A_2B_1A_1A_1A_1)$ contrasts with the predicted sequence $(B_2A_2A_2B_1A_1A_1A_1|B_1A_1A_1B_2A_2A_2A_2)$ from the MFD.
}
\label{fig:610810}
\end{figure*}

Thanks to the computational framework proposed in Sec.\,\ref{subsec:framework}, high-$\Sigma$ CSL GBs can be accurately calculated.
We verify Farey fractions $2/12$ and $2/16$, corresponding to $\Sigma 37~(610)$ and $\Sigma 65~(810)$ GBs, respectively, as shown in \cref{fig:610810}.
In \cref{fig:610810}(a), since $2/12 = 1/5 \boxplus 1/7$, the predicted sequence through MFD is $(B_2A_2B_1A_1A_1|B_1A_1B_2A_2A_2)$, with $A_1$ count of $2$ and $1$ in two parts.
Furthermore, theoretical calculation reveals that $\Sigma 37~(610)$ GB has the $(B_2A_2B_1A_1A_1)$ periodic arrangement, which is half of the predicted sequence where the number of $A_1$ is larger.
Similarly, in \cref{fig:610810}(b), $2/16 = 1/7 \boxplus 1/9 = 1/7 \boxplus 1/7 \boxplus 0/2$, the MFD predicts that the periodic arrangement of $\Sigma 65~(810)$ GB is $(B_2A_2A_2B_1A_1A_1A_1|B_1A_1A_1B_2A_2A_2A_2)$.
The theoretically calculated SU periodic sequence is $(B_2A_2A_2B_1A_1A_1A_1)$, also consistent with the arrangement rule proposed above.
The predicted rule developed for A15 GBs is different from the previous work for FCC GBs, where the predicted sequence of MFD does not require reduction \cite{inoue20213d}.

\subsubsection{Predicting GBs for $\theta \in (36.9^\circ,90^\circ)$}

\begin{figure*}[!hbpt]
\centering
 \includegraphics[width=0.8\textwidth]{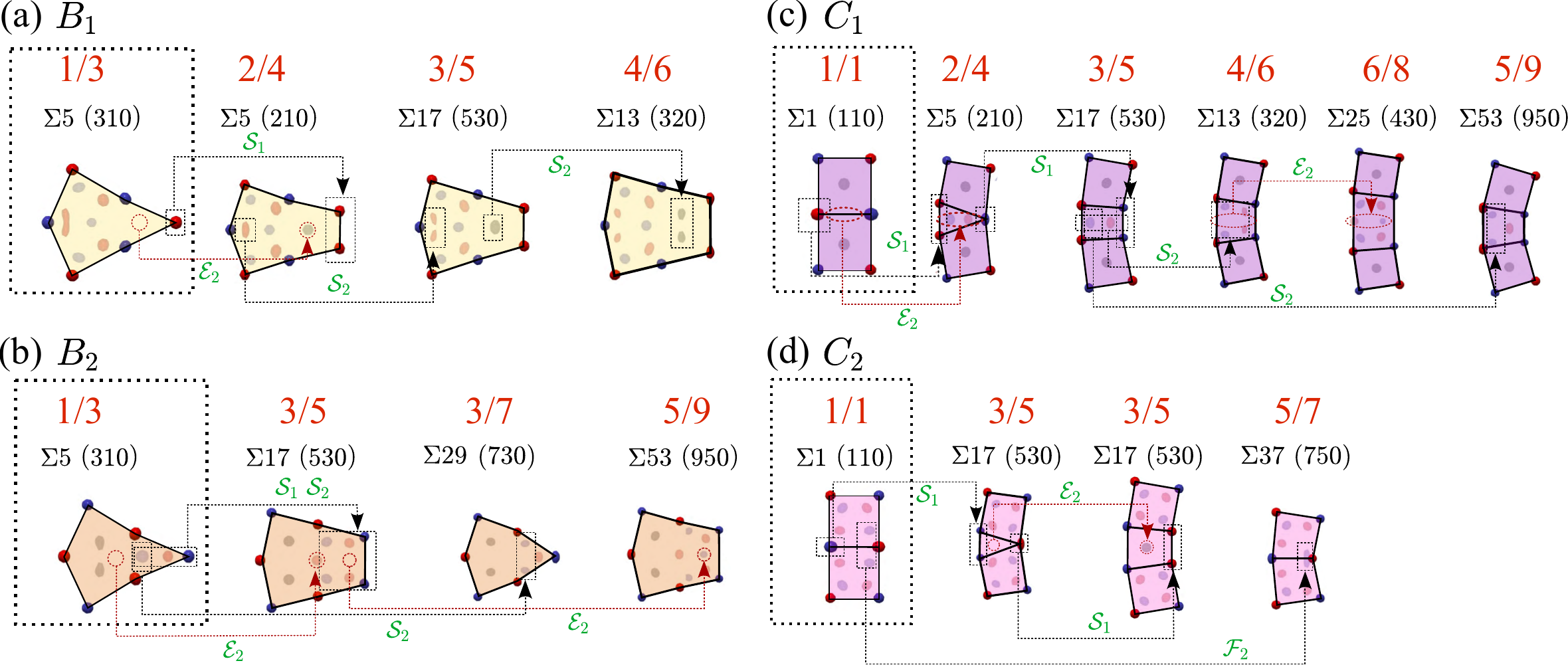}
\caption{
Shapes and deformations of SUs $B_{1}$, $B_{2}$, $C_{1}$, and $C_{2}$ in GBs with $36.9^{\circ}$-$90^{\circ}$ misorientation angles within MFD in \cref{fig:farey}.
SUs are ordered by ascending Farey denominators within each category.
Where identical SUs appear across multiple GBs, we retain the instance with the smallest Farey denominator, removing duplicates.
Dashed boxes indicate the basic SUs in delimiting GBs.
Green symbols denote deformation types of SUs, where subscripts ``1" and ``2" represent lattice spheres and decorative spheres, respectively.
SU deformation modes include splitting of lattice spheres ($\mathcal{S}_{1}$), splitting of decorative spheres ($\mathcal{S}_{2}$), emergence of new decorative spheres ($\mathcal{E}_{2}$), and fusion of adjacent decorative spheres ($\mathcal{F}_{2}$).}
\label{fig:SU2}
\end{figure*}

\begin{figure*}[!hbpt]
\centering
 \includegraphics[width=0.8\textwidth]{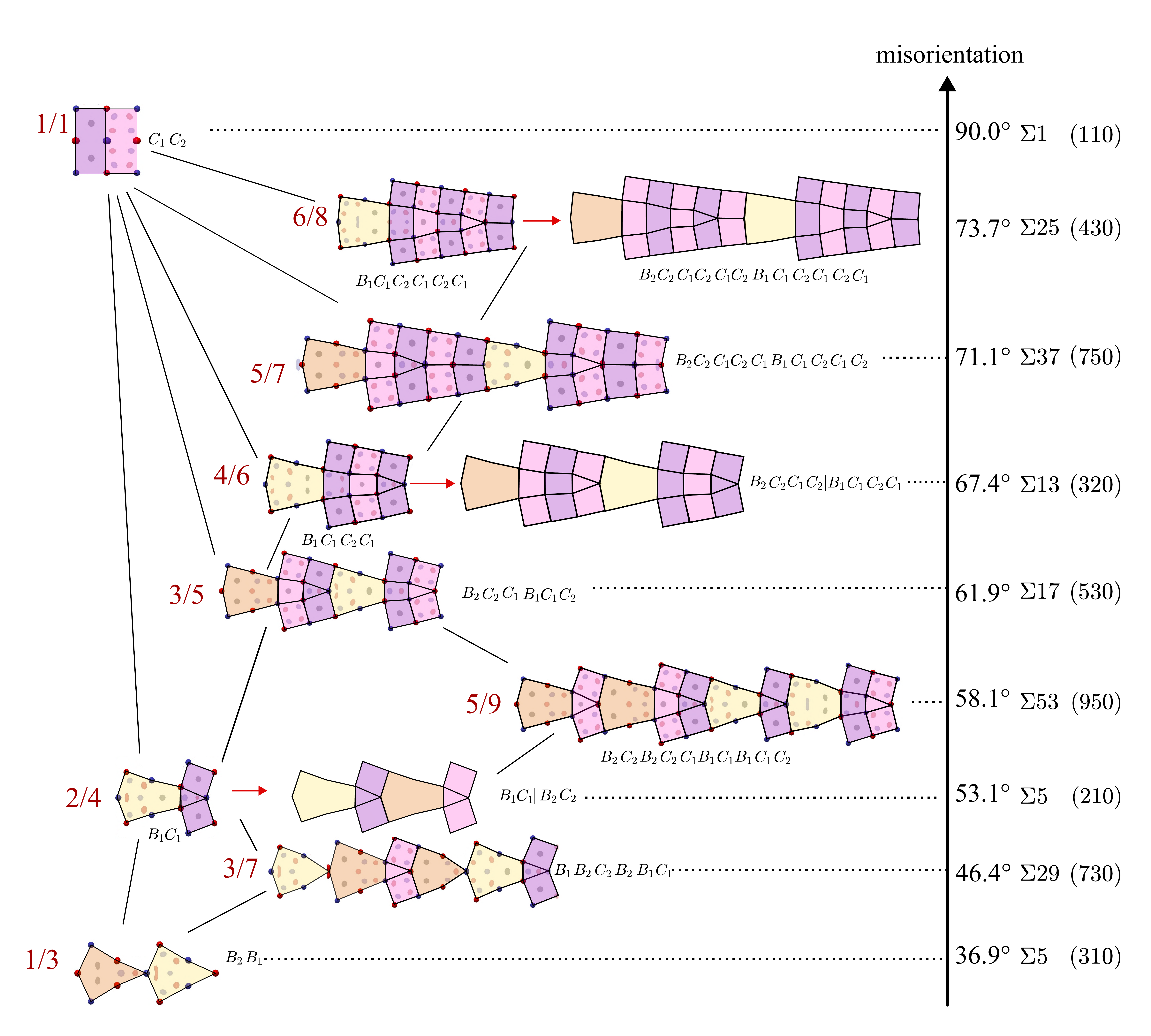}
\caption{
Structural prediction of $[001]$ CSL STGBs in A15 with misorientation angles from $36.9^\circ$ to $90^\circ$ using MFD and theoretical calculations.
An SU binary tree is constructed based on MFD in \cref{fig:farey} by positioning GBs near corresponding Farey fractions.
For irreducible Farey fractions (e.g., $3/5$, $5/7$), the periodic sequences of SUs from theoretical calculations match MFD predictions.
For reducible fractions (e.g., $4/6$, $6/8$), both theoretical GB structures and MFD predictions are displayed.
}
\label{fig:farey2}
\end{figure*}

After exploring SU arrangement rules in GBs with misorientation angles in the range of $0^\circ$ to $36.9^\circ$, we now focus on the range of $36.9^\circ$ to $90^\circ$.
GBs in this range are composed of SUs $B$ and $C$, and corresponding Farey fractions range from $1/3$ to $1$.
\cref{fig:SU2} shows concrete shapes of SUs $B_1$, $B_2$, $C_1$, and $C_2$ within delimiting GBs and several low-$\Sigma$ CSL GBs.
Dashed boxes contain basic shapes of SUs in delimiting GBs.
Compared with deformation types in \cref{fig:SU1}, SUs lack $\mathcal{D}_2$ but exhibit a new deformation where two decorative spheres fuse into one ($\mathcal{F}_2$).
The $\mathcal{S}_1$ deformation of SUs of type $B$ in \cref{fig:SU1}(c)(d) occurs at the leftmost lattice sphere, while in \cref{fig:SU2}, it occurs at the rightmost lattice sphere, resulting in different shapes of the SU $B$ across the two ranges.
The deformation process of SU $B_1$ following Farey fractions is $1/3\xrightarrow[  ]{\mathcal{S}_1\mathcal{E}_2}2/4\xrightarrow[]{\mathcal{S}_2}3/5\xrightarrow[]{\mathcal{S}_2}4/6$, that of $B_2$ is $1/3\xrightarrow[]{\mathcal{S}_1\mathcal{S}_2\mathcal{E}_2}3/5\xrightarrow[]{\mathcal{E}_2}5/9$ and $1/3\xrightarrow[]{\mathcal{S}_2}3/7$, that of $C_1$ is $1/1\xrightarrow[]{\mathcal{S}_1\mathcal{E}_2}2/4\xrightarrow[  ]{\mathcal{S}_1}3/5\xrightarrow[]{\mathcal{S}_2}4/6\xrightarrow[]{\mathcal{E}_2}6/8$ and $3/5\xrightarrow[]{\mathcal{S}_2}5/9$, and that of $C_2$ is $1/15\xrightarrow[]{\mathcal{S}_1}3/5\xrightarrow[]{\mathcal{S}_1\mathcal{E}_2}6/8$ and $1/15\xrightarrow[]{\mathcal{F}_2}5/7$.
SU classification remains unaffected by deformations.

\cref{fig:farey2} shows GBs corresponding to Farey fractions between $1/3$ and $1$ in modified Farey sequence $F^\prime_9$.
SU periodic sequences of these GBs are composed of $(B_1 C_1 C_2 C_1 \dots)$ and $(B_2 C_2 C_1 C_2 \dots)$, where the number of type $C$ may be $0$, otherwise the last symbol is either $C_1$ or $C_2$.
Farey fractions are also classified as irreducible or reducible.
For irreducible fractions, SU periodic sequences can be directly deduced from MFD.
For example, $3/5 = 2/4 \boxplus 1/1 $ implies that SU periodic sequence of $\Sigma 17~(530)$ GB is composed of $(B_1C_1B_2C_2)$ in $\Sigma 5~(210)$ GB and $(C_1C_2)$ in $\Sigma 1~(110)$ GB. 
For reducible fractions, the MFD-predicted sequence requires reduction.
The Farey fraction corresponding to $\Sigma 5~(210)$ GB is $2/4 = 1/3 \boxplus 1/1$.
The SU periodic sequence predicted by the MFD is $(B_1C_1|B_2C_2)$, while the theoretical calculation result is $(B_1C_1)$, which retains half of the predicted sequence that contains $B_1$.
High-$\Sigma$ CSL GBs such as $\Sigma 13~(320)$ and $\Sigma 25~(430)$ GBs have SU periodic sequences $(B_1C_1C_2C_1)$ and $(B_1C_1C_2C_1C_2C_1)$, respectively, which also follow this reduction rule.

\section{Conclusions}
\label{sec:conclusion}

In this work, we establish a theoretical approach incorporating MFD and computational framework to predict tilt A15 CSL GBs, particularly high-$\Sigma$ CSL ones.
We investigate $[001]$ A15 CSL STGBs with misorientation angles from $0^\circ$ to $90^\circ$.
SUs $A_1$, $A_2$, $B_1$, $B_2$, $C_1$, and $C_2$ are identified, along with three delimiting GBs: $\Sigma 1~(100)~(0^{\circ})$, $\Sigma 5~(310)~(36.9^{\circ})$, and $\Sigma 1~(110)~(90.0^{\circ})$.
Dividing the misorientation range into two intervals ($0^\circ$-$36.9^\circ$ and $36.9^\circ$-$90^\circ$), we predict SU configurations using the MFD.
The computational framework is then applied to numerical simulations of A15 $[001]$ CSL STGBs to determine specific arrangements of SUs.
When Farey fractions are irreducible, the MFD's predictions match theoretical calculations of GB structures.
For reducible Farey fractions, the prediction sequence requires reduction to align with theoretical SU periodicity.
Specifically, in the $0^\circ$-$36.9^\circ$ misorientation range, the theoretical sequence retains the subsequence with more $A_1$ SUs.
In the $36.9^\circ$-$90^\circ$ range, the subsequence containing $B_1$ SU remains.
This work establishes a universal methodology for predicting GB structures in A15 materials, providing theoretical guidance for microstructure design.

\section*{Acknowledgments}
This work is partially supported by 
the National Key R\&D Program of China (2023YFA1008802), 
the National Natural Science Foundation of China (12171412), 
the Science and Technology Innovation Program of Hunan Province (2024RC1052), 
the Innovative Research Group Project of Natural Science Foundation of Hunan Province of China (2024JJ1008),
the Postgraduate Scientific Research Innovation Project of Hunan Province (CX20230634), 
and the High Performance Computing Platform of Xiangtan University.

\bibliography{ref}

\end{document}